\documentclass{IEEEtran4PSCC}

\usepackage{cite}

\ifCLASSINFOpdf
   \usepackage[pdftex]{graphicx}
\else
   \usepackage[dvips]{graphicx}
\fi
\usepackage[tbtags]{amsmath}
\ifCLASSOPTIONcompsoc
  \usepackage[caption=false,font=normalsize,labelfont=sf,textfont=sf]{subfig}
\else
  \usepackage[caption=false,font=footnotesize]{subfig}
\fi
\usepackage{comment}
\usepackage{mathtools}
\usepackage{tikz}
\usepackage{xcolor}
\usepackage{dblfloatfix}
\usepackage[]{footmisc}
\usepackage[normalem]{ulem}
\usepackage{multirow}
\usepackage{siunitx}
\usepackage{etoolbox}
\allowdisplaybreaks

\usepackage{cleveref}
\crefname{equation}{equation}{equations}
\Crefname{equation}{Equation}{Equations}
\crefrangelabelformat{equation}{(#3#1#4--#5#2#6)}

\crefmultiformat{equation}{equations (#2#1#3}{, #2#1#3)}{#2#1#3}{#2#1#3}
\Crefmultiformat{equation}{Equations (#2#1#3}{, #2#1#3)}{#2#1#3}{#2#1#3}

\usepackage[justification=justified,font=footnotesize,skip=0pt]{caption}
\makeatletter
\let\old@ps@headings\ps@headings
\let\old@ps@IEEEtitlepagestyle\ps@IEEEtitlepagestyle
\def\psccfooter#1{%
    \def\ps@headings{%
        \old@ps@headings%
        \def\@oddfoot{\strut\hfill#1\hfill\strut}%
        \def\@evenfoot{\strut\hfill#1\hfill\strut}%
    }%
    \def\ps@IEEEtitlepagestyle{%
        \old@ps@IEEEtitlepagestyle%
        \def\@oddfoot{\strut\hfill#1\hfill\strut}%
        \def\@evenfoot{\strut\hfill#1\hfill\strut}%
    }%
    \ps@headings%
}
\makeatother

\begin{document}
%
\title{Practical Considerations of DER Coordination with Distributed Optimal Power Flow}

\author{
\IEEEauthorblockN{Daniel Gebbran}
\IEEEauthorblockA{University of Sydney \\
Sydney, Australia\\
daniel.gebbran@sydney.edu.au\\
}
\vspace{2mm}
\IEEEauthorblockN{Wibowo Hardjawana}
\IEEEauthorblockA{University of Sydney \\
Sydney, Australia\\
wibowo.hardjawana@sydney.edu.au\\
}
\and
\IEEEauthorblockN{Sleiman Mhanna}
\IEEEauthorblockA{University of Melbourne\\
Melbourne, Australia\\
sleiman.mhanna@unimelb.edu.au
}
\vspace{2mm}
\IEEEauthorblockN{Branka Vucetic}
\IEEEauthorblockA{University of Sydney \\
Sydney, Australia\\
branka.vucetic@sydney.edu.au\\
}
\and 
\IEEEauthorblockN{Archie C. Chapman}
\IEEEauthorblockA{University of Queensland\\ 
Brisbane, Australia\\
archie.chapman@uq.edu.au
}
\vspace{2mm}
\IEEEauthorblockN{Gregor Verbi\v{c}}
\IEEEauthorblockA{University of Sydney \\
Sydney, Australia\\
gregor.verbic@sydney.edu.au\\
}


}

\makeatletter
\patchcmd{\@maketitle}
  {\addvspace{0.5\baselineskip}\egroup}
  {\addvspace{-1.5\baselineskip}\egroup}
  {}
  {}
\makeatother

\maketitle

\begin{abstract}
The coordination of prosumer-owned, behind-the-meter distributed energy resources (DER) can be achieved using a multiperiod, distributed optimal power flow (DOPF), which satisfies network constraints and preserves the privacy of prosumers. 
To solve the problem in a distributed fashion, it is decomposed and solved using the alternating direction method of multipliers (ADMM), which may require many iterations between prosumers and the central entity (i.e., an aggregator). Furthermore, the computational burden is shared among the agents with different processing capacities. Therefore, computational constraints and communication requirements may make the DOPF infeasible or impractical. 
In this paper, part of the DOPF (some of the prosumer subproblems) is executed on a Raspberry Pi-based hardware prototype, which emulates a low processing power, edge computing device. 
Four important aspects are analyzed using test cases of different complexities. 
The first is the computation cost of executing the subproblems in the edge computing device. 
The second is the algorithm operation on congested electrical networks, which impacts the convergence speed of DOPF solutions. 
Third, the precision of the computed solution, including the trade-off between solution quality and the number of iterations, is examined. 
Fourth, the communication requirements for implementation across different
communication networks are investigated. 
The above metrics are analyzed in four scenarios involving 26-bus and 51-bus networks.
\end{abstract}

\begin{IEEEkeywords}
Distributed optimal power flow (DOPF), distributed energy resources (DER), ADMM, prosumers, demand response, communication latency, edge computing.
\end{IEEEkeywords}

\thanksto{\noindent Corresponding author: daniel.gebbran@sydney.edu.au. }

% ...

\vspace{-6mm}

\section{Introduction}
Decentralizing power systems by integrating distributed energy resources (DER) at the prosumer level offers economic and technical benefits for both owners and network operators, but requires careful coordination to minimize negative impacts on the grid \cite{OpenEnergyNetworks}.
In this context, distributed optimal power flow (DOPF) methods have been shown to successfully coordinate DER \cite{Scott2019, Andrianesis2019, Attarha2020, Guerrero_2020}, ensuring network constraints are always satisfied whilst also preserving prosumer privacy and prerogatives. 
However, there is currently limited literature analyzing practical applications of DOPF \cite{Scott2019,Gebbran_AAMAS}, and important implementation aspects have not been discussed in sufficient detail, such as: (i) the solution time of DOPF on actual distributed hardware, (ii) operation of the algorithm in congested electrical networks, (iii) the precision of the solution and (iv) the communication requirements for implementation, such as latency 
requirements. 
%
To fill this gap in the literature, in this paper we present a DPOF deployment on edge computing devices, and discuss its characteristics and real-world performance.
%

\vspace{-2mm}
\subsection{Background}

The AC optimal power flow problem is typically solved using interior point methods, because it is a nonconvex problem. Although these methods cannot guarantee global optimality in general (since they solve to local optimality), the resulting solution is guaranteed to be feasible. 
However, the OPF quickly becomes intractable when considering DER due to the sheer number of variables involved. This motivates investigations into distributed approaches, of which several methods have been applied: dual decomposition, analytic target cascading, auxiliary problem principle, optimality condition decomposition, gradient dynamics, dynamic programming with message passing, and the alternating direction method of multipliers (ADMM). A comprehensive review of their implementations can be found in \cite{Molzahn2017}. 

ADMM \cite{Boyd2011} has been widely used to solve large-scale OPF problems \cite{Molzahn2017}, as it allows for for flexible decompositions of the original OPF problem. They range from network sub-regions \cite{Kim1997} down to an element-wise (e.g., generators, buses, and lines) decomposition \cite{Mhanna2019}. In ADMM, each of the resulting decomposed parts solves a subproblem and exchanges messages with a central aggregator (or between other agents) until convergence is achieved \cite{Guo_2018}. 
A decomposition at the point of connection between prosumers and the network was deemed a pratical balance for DER coordination
\cite{Scott2019, Andrianesis2019, Attarha2020}. It preserves privacy of prosumers and allows for parallelization of subproblems (benefits against centralized approaches), and offers quicker solutions (smaller number of iterations) when compared to fully decentralized approaches. 
This approach has been demonstrated to successfully coordinate DER in real-world scenarios in a recent Australian trial \cite{Scott2019}, and can be implemented on edge computing devices (at individual prosumers), benefiting from subproblem parallelization to distribute the computational load \cite{Gebbran_AAMAS}. 

Because this approach is very recent, there is sparse literature and a dearth of information regarding practical considerations for this DER coordination method. 

\subsection{Contributions}
This work offers important technical insights into modeling and deploying DER coordination methods using DOPF. To offer a solid testbed, part of the subproblems is deployed on a hardware prototype, based on Raspberry Pis 3B+ (RPis) -- a small, single-board computer. This allows for a more realistic analysis, emulating an edge computing archetype where prosumer computations are conducted on embedded hardware. The remainder of the problem is solved on a PC. Four different test cases are simulated, involving two networks and two time horizons, which allows for comparison across different setups. 
The paper focuses on four principal characteristics of the problem, which can be summarized in the following contributions:

\begin{itemize}
    \item Quantification of computation times for the DOPF implemented across edge computing devices.
    \item Investigation of algorithm execution on normal operation versus congested system conditions.
    \item Analysis of solution precision, including trade-offs between solution quality and computational burden.
    \item Discussion of communication requirements for implementation on modern communication networks.
\end{itemize}

\subsection{Paper Structure}

The remainder of the paper is structured as follows: Section II formulates the DOPF, including the initial problem, the decomposition and the resulting distributed problem formulation. Section III discusses details of the implementation, including algorithm specifications, hardware description and details of the test networks. Section IV presents the results and discusses each of the four main proposed metrics. Finally, Section V presents a general discussion on the results and Section VI finishes with concluding remarks. 


\vspace{-1mm}
\section{MOPF Formulation}

The proposed approach for DER coordination is formulated as a multi-period optimal power flow (OPF) problem. It consists of two levels. At the lower level, prosumers \emph{schedule} their DER, minimizing energy expenditure\footnote{When ${c}^{\text{tou}} \! > \! {c}^{\text{fit}}$, as is the case in Australia, this corresponds to PV self-consumption.}. At the upper level, the distribution network system operator (DNSP) \emph{coordinates} prosumers’ actions to minimize the network objective - whilst abiding by network limits and operational constraints.

The objective function of this problem is:
\begin{equation} \label{eq:Central_OPF_1st}
\begin{split}
&\underset{\boldsymbol{x}, \: \boldsymbol{z}} {\mbox{minimize}} \quad F(\boldsymbol{x},\boldsymbol{z}) \coloneqq f(\boldsymbol{x}) + \sum_{h\in \mathcal{H}} g_h(\boldsymbol{z}_h) \\
& = \underset{t \: \in \: \mathcal{T}}{{\sum}} \Big( {c}_{2}{ ( {p}_{g,t}^{+}) }^2 + {c}_{1} {p}_{g,t}^{+} + {c}_{0} + \underset{h \: \in \: \mathcal{H}}{\sum} \big( {c}^{\text{tou}}_{i}{ {p}_{h,t}^{+} } - {c}^{\text{fit}}{ {p}_{h,t}^{-} \big) } \Big) , 
\end{split}
\end{equation}
\normalsize
where $f(\boldsymbol{x})$ represents the network OPF objective function (which can include, for example, loss minimization, peak load reduction or minimizing the use of backup diesel as in \cite{Scott2019}), 
$g_{h}(\boldsymbol{z}_{h})$ are prosumer objective functions for each household $h$, with a fixed time-of-use tariff for purchasing energy, and a feed-in-tariff for selling energy, 
$\mathcal{H}$ is the set of prosumers, 
$\boldsymbol{x}$ is the set of network variables (active/reactive power flows, and voltages, for each $t \in \mathcal{T}$), 
and $\boldsymbol{z}_{h}$ is the set of internal variables of prosumer $h$ for each $t \in \mathcal{T}$ (e.g., battery power flows), 
which compose the set of variables for all prosumers $\boldsymbol{z} \coloneqq {\{ \boldsymbol{z}_h\}}_{h\in \mathcal{H}}$. 

The network constraints for a single-phase OPF are shown below.\footnote{A balanced three-phase network is assumed for simplicity. It can be modeled as a single phase. However, the single-phase model can be readily extended, e.g. including unbalanced networks with a combination of single- and three-phase connections \cite{Scott2019}, increasing the formulation's complexity.} They are given for each bus $i \in \mathcal{B}$, and for each time interval $t \in \mathcal{T}$:
%
\begin{subequations} \label{eq:OPF_Constraints}
\begin{align}
    & {p}_{g,t} - {p}_{h,t} = v_{i,t} \underset{j \: \in \: \mathcal{B}}{\sum} v_{j,t}(g_{ij}\cos\theta_{ij,t}+b_{ij}\sin\theta_{ij,t}), \label{eq:OPF_P-flow}\\
    & {q}_{g,t} - {q}_{h,t} = v_{i,t} \underset{j \: \in \: \mathcal{B}}{\sum} v_{j,t}(g_{ij}\sin\theta_{ij,t}-b_{ij}\cos\theta_{ij,t}), \label{eq:OPF_Q-flow}\\
    & v_{r,t} = 1, \quad \theta_{r,t} = 0, \label{eq:OPF_V-slack}\\
    & \underline{v}_{i} \leq v_{i,t} \leq \overline{v}_{i}, \label{eq:OPF_V-lim}\\
    %
    %
    & \underline{p}_{g,t} \leq {p}_{g,t} \leq \overline{p}_{g,t}, \quad %
    \underline{q}_{g,t} \leq {q}_{g,t} \leq \overline{q}_{g,t}, \label{eq:OPF_G-limits}
\end{align}
\normalsize
\end{subequations}
where ${p}_{g,t}, {q}_{g,t}$ are the total net active/reactive power from the reference bus, 
${p}_{h,t}, {q}_{h,t}$ are the total net active/reactive power to prosumer $h$ connected to bus $i$, and $\theta_{ij,t} = \theta_{i,t} - \theta_{j,t}$ is the angle difference between bus $i$ and its neighboring bus $j$. Additionally, \eqref{eq:OPF_P-flow}, \eqref{eq:OPF_Q-flow} model the power flow equations, (\ref{eq:OPF_V-slack}) models the reference, and \eqref{eq:OPF_V-lim}, \eqref{eq:OPF_G-limits} represent voltage and generator (lower and upper) limits. 
Moreover, let ${p}_{h,t} = {p}_{h,t}^{+} - {p}_{h,t}^{-}$ be composed of the non-negative terms ${p}_{h,t}^{+}, {p}_{h,t}^{-}$, representing imported and exported power. The same applies for ${p}_{g,t}$.\footnote{Note that because the second term in \eqref{eq:Central_OPF_1st} is a convex piecewise linear function, \emph{at least one} of the variables ${p}_{h,t}^{+}$ and ${p}_{h,t}^{-}$ can be zero at time slot $t$. This therefore obviates the need to use binary variables.}

Each prosumer $h \in \mathcal{H}$ is subject to its own constraints. The equation modeling the power balance is, $ \forall \: t \in \mathcal{T}, h \in \mathcal{H}$: 
%
\begin{equation} \label{eq:HEM_Balance}
{p}_{h,t} = {p}_{h,t}^{\text{bat}} + {p}_{h,t}^{\text{d}} - {p}_{h,t}^{\text{PV}},
\end{equation}
\normalsize
where ${p}_{h,t}$ is the total net power (exchanged with the grid) of household $h$, with $\underline{p}_{h,t} \leq {p}_{h,t} \leq \overline{p}_{h,t}$, ${p}_{h,t}^{\text{bat}}$ is the scheduled battery charging power, with $\underline{p}_{h,t}^{\text{bat}} \leq {p}_{h,t}^{\text{bat}} \leq \overline{p}_{h,t}^{\text{bat}}$; ${p}_{h,t}^{\text{d}}$ is the household non-controllable (fixed) demand, and ${p}_{h,t}^{\text{PV}}$ is the PV generation power output, which can be curtailed if necessary (the total available PV power is $\tilde{p}_{h,t}^{\text{PV}} \geq {p}_{h,t}^{\text{PV}} \geq 0$). 

The battery constraints are, $ \forall \: t \in \mathcal{T}, h \in \mathcal{H}$:
%
\begin{subequations} \label{eq:HEM_Bat}
\begin{align}
& {p}_{h,t}^{\text{bat}} = {p}_{h,t}^{\text{ch}} - {p}_{h,t}^{\text{dis}}, \label{eq:HEM_bat_Dec}\\
& {SoC}_{h,0} \le {SoC}_{h,T}, \label{eq:HEM_SoC0}\\
& {SoC}_{h,t} = {SoC}_{h, t - \Delta t } + (\eta_{h}^{\text{ch}} {p}_{h,t}^{\text{ch}} - {{p}_{h,t}^{\text{dis}}}/{\eta_{h}^{\text{dis}}})\Delta t, \label{eq:HEM_bat_SoC}
\end{align}
\end{subequations}
\normalsize
where ${p}_{h,t}^{\text{ch}}, {p}_{h,t}^{\text{dis}} \geq 0$ compose the battery charging/discharging power; ${SoC}_{h,t}$ is the battery state-of-charge, with $\underline{SoC}_{h,t} \leq {SoC}_{h,t} \leq \overline{SoC}_{h,t}$\footnote{Including (\ref{eq:HEM_SoC0}) avoids full battery depletion - without considering the next time horizon. Replacing it is recommended for algorithm implementation using a rolling horizon basis.}, $\eta_{h}$ is the battery charge or discharge efficiency, and $\Delta t$ is the time interval within $\mathcal{T}$.

To rewrite the problem in its compact form, let the network constraints (\ref{eq:OPF_Constraints}) define a feasible set $\mathcal{X}$ for the network variables $\boldsymbol{x}$ and prosumer constraints \eqref{eq:HEM_Balance}, \eqref{eq:HEM_Bat} define a feasible set $\mathcal{Z}_{h}$ for the variables $\boldsymbol{z}_h$ 
of each prosumer $h \in \mathcal{H}$. 
Henceforth, $\boldsymbol{x} \in \mathcal{X}$ and $\boldsymbol{z}_h \in \mathcal{Z}_{h}$, with $\boldsymbol{z} \in \mathcal{Z}$ (the feasible set for all prosumer variables). We can now write:
%
\begin{equation} \label{eq:Central_OPF_2nd}
\underset{\boldsymbol{x} \in \mathcal{X}, \: \boldsymbol{z} \in \mathcal{Z}} {\mbox{minimize}} \quad F(\boldsymbol{x},\boldsymbol{z})
\end{equation}

Two problems arise if we are to solve this MOPF centrally. First, the privacy of all prosumers is violated, since all data (battery information, consumption data, etc) for each house has to be sent to the central computing entity. 
Second, the problem is computationally hard because it consists of a non-convex network problem \cite{Verma2019}. Solving such a large-scale nonlinear problem is extremely challenging, especially given a potentially large number (several tens or even hundreds) of prosumer subproblems. 
Hence, a distributed approach is applied to solve this MOPF with DR problem.

\vspace{-1mm}
\subsection{Decomposed Model}
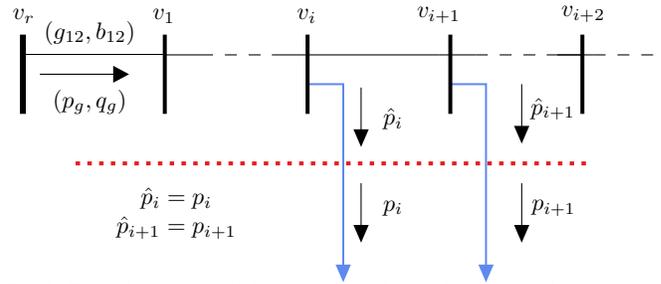
\begin{figure}[!t]
\begin{tikzpicture}[x=0.75pt,y=0.75pt,yscale=-1,xscale=0.9]

\draw [color={rgb, 255:red, 94; green, 137; blue, 240 } ,draw opacity=1 ][line width=0.75]  (190,148) -- (190,50) -- (170,50) ;

\draw [shift={(190,150)}, rotate = 270] [fill={rgb, 255:red, 94; green, 137; blue, 240 } ,fill opacity=1 ][line width=0.75] [draw opacity=0] (8.93,-4.29) -- (0,0) -- (8.93,4.29) -- cycle  ;
\draw [color={rgb, 255:red, 94; green, 137; blue, 240 } ,draw opacity=1 ][line width=0.75]  (270,148) -- (270,50) -- (250,50) ;

\draw [shift={(270,150)}, rotate = 270] [fill={rgb, 255:red, 94; green, 137; blue, 240 } ,fill opacity=1 ][line width=0.75] [draw opacity=0] (8.93,-4.29) -- (0,0) -- (8.93,4.29) -- cycle  ;
\draw [line width=2.25]  (10.5,25) -- (10.5,65) ;

\draw  (10,35) -- (90,35) ;

\draw [line width=1.5]  (90,25) -- (90,65) ;

\draw  (20,45) -- (68,45) ;
\draw [shift={(70,45)}, rotate = 180] [fill={rgb, 255:red, 0; green, 0; blue, 0 } ][line width=0.75] [draw opacity=0] (8.93,-4.29) -- (0,0) -- (8.93,4.29) -- cycle  ;

\draw [dash pattern={on 4.5pt off 4.5pt}] (110.33,35.22) -- (150,35.22) ;

\draw [line width=1.5]  (170,25) -- (170,65) ;

\draw  (150,35.22) -- (169.5,35.22) ;

\draw [line width=1.5]  (250,25) -- (250,65) ;

\draw  (169.5,35.22) -- (244,35.22) ;

\draw [dash pattern={on 4.5pt off 4.5pt}] (264.33,35.22) -- (304,35.22) ;

\draw  (304,35.22) -- (323.5,35.22) ;

\draw  (244,35.22) -- (264.33,35.22) ;

\draw [line width=1.5]  (324,25) -- (324,65) ;

\draw [color={rgb, 255:red, 229; green, 16; blue, 19 } ,draw opacity=0.95 ][line width=1.5] [dash pattern={on 1.69pt off 2.76pt}] (40,90) -- (330,90) ;

\draw  (200,51.71) -- (200,70) -- (200,79.71) ;
\draw [shift={(200,81.71)}, rotate = 270] [fill={rgb, 255:red, 0; green, 0; blue, 0 } ][line width=0.75] [draw opacity=0] (8.93,-4.29) -- (0,0) -- (8.93,4.29) -- cycle  ;

\draw  (200,100) -- (200,128) ;
\draw [shift={(200,130)}, rotate = 270] [fill={rgb, 255:red, 0; green, 0; blue, 0 } ][line width=0.75] [draw opacity=0] (8.93,-4.29) -- (0,0) -- (8.93,4.29) -- cycle  ;

\draw  (290,100) -- (290,128) ;
\draw [shift={(290,130)}, rotate = 270] [fill={rgb, 255:red, 0; green, 0; blue, 0 } ][line width=0.75] [draw opacity=0] (8.93,-4.29) -- (0,0) -- (8.93,4.29) -- cycle  ;

\draw  (290,50) -- (290,78) ;
\draw [shift={(290,80)}, rotate = 270] [fill={rgb, 255:red, 0; green, 0; blue, 0 } ][line width=0.75] [draw opacity=0] (8.93,-4.29) -- (0,0) -- (8.93,4.29) -- cycle  ;

\draw  (90,35.22) -- (110.33,35.22) ;

\draw [dash pattern={on 4.5pt off 4.5pt}] (330.33,35.22) -- (370,35.22) ;

\draw  (310,35.22) -- (330.33,35.22) ;

\draw (48,59) node [scale=0.9] {$(p_{g},q_{g})$};
\draw (11,15.5) node [scale=0.9] {$v_{r}$};
\draw (89.5,15.5) node [scale=0.9] {$v_{1}$};
\draw (169.5,15.5) node [scale=0.9] {$v_{i}$};
\draw (243.5,15) node [scale=0.9] {$v_{i+1}$};
\draw (324.5,14.5) node [scale=0.9] {$v_{i+2}$};
\draw (218,67.5) node [scale=0.9] {$\hat{p}_{i}$};
\draw (48,24) node [scale=0.9] {$(g_{12},b_{12})$};
\draw (307.5,62.5) node [scale=0.9] {$\hat{p}_{i+1}$};
\draw (218,112.5) node [scale=0.9] {$p_{i}$};
\draw (308,112.5) node [scale=0.9] {$p_{i+1}$};
\draw (96.5,107.5) node [scale=0.9] {$\hat{p}_{i} =p_{i}$};
\draw (96.5,122.5) node [scale=0.9] {$\hat{p}_{i+1} =p_{i+1}$};
\end{tikzpicture}
\caption{Example of network decomposition depicted over a single time period by duplication of coupling variables.}
\label{fig:Network_Decomposition}
\end{figure}

Normally, we would not be able to solve (\ref{eq:Central_OPF_2nd}) in a distributed fashion. This is because the variables corresponding to the prosumer power consumption appear in both $\mathcal{X}$ and $\mathcal{Z}$. To enable a decomposable structure for the problem, we create two copies of all prosumer power profiles, as shown in Fig. \ref{fig:Network_Decomposition}, introducing the following coupling constraints:
\begin{equation} \label{eq:Coupling}
\hat{p}_{h,t} = {p}_{h,t}, \qquad \forall \; h \in \mathcal{H}, \; t \in \mathcal{T},%
\end{equation}
where the left-hand term is a copy for the network problem, $\hat{p}_{h,t} \in \mathcal{X}$, and the right-hand term is a copy for the prosumer problem, ${p}_{h,t} \in \mathcal{Z}_{h}$. 


Now, we can treat prosumer subproblems separately from the network, coupled only through prosumer power consumption. 
Problem (\ref{eq:Central_OPF_2nd}) can now be decomposed because $f(\boldsymbol{x})$ and $g_{h}(\boldsymbol{z}_{h})$ are themselves separable. 
In more detail, duplicating the variables as (\ref{eq:Coupling}) enables us to rewrite (\ref{eq:Central_OPF_2nd}) as:
\begin{subequations} \label{eq:DecomposedModel}
\begin{align}
&\underset{\boldsymbol{\hat{x}} \in \mathcal{\hat{X}}, \: \boldsymbol{{z}} \in \mathcal{{Z}}} {\mbox{minimize}} \quad F(\boldsymbol{\hat{x}},\boldsymbol{z}), \label{eq:Decomposed0}\\
& \text{subject to:} \hspace{0.55cm} \text{(\ref{eq:Coupling})} \label{eq:Decomposed1},%
\end{align}
\end{subequations}
where $\boldsymbol{\hat{x}}$ is the original set of problem variables with the addition of the network copy of prosumer's power profiles (\ref{eq:Coupling}), and $\mathcal{\hat{X}}$ is the new feasible region of the network problem. Now, the sets of variables $\mathcal{\hat{X}}$ and $\mathcal{Z}$ are decoupled, and (\ref{eq:Decomposed0}) is separable if \eqref{eq:Decomposed1} is relaxed. The resulting decoupled problem is illustrated in Fig. \ref{fig:Network_Decomposition}. We will exploit this structure to solve (\ref{eq:DecomposedModel}) in a distributed fashion.

Finally, we write the augmented (partial) Lagrange function:
\begin{equation} \label{eq:Lagrangian}
\begin{split}
L & := f(\boldsymbol{\hat{x}}) +
\underset{h \in \mathcal{H}}{{\sum}} \Big( g_h(\boldsymbol{z}_h) + \underset{t \in \mathcal{T}}{{\sum}}
\big( \frac{\rho}{2} (\hat{p}_{h,t} - {p}_{h,t})^2 \\
& \hspace{1mm} + \lambda_{h,t} (\hat{p}_{h,t} - {p}_{h,t}) \big) \Big) = F(\boldsymbol{\hat{x}},\boldsymbol{z}) + \underset{h \in \mathcal{H}}{{\sum}} L_h,
\end{split}
\end{equation}
\normalsize
where $\rho$ is a penalty parameter and ${\lambda}_{h,t}$ is the dual variable associated with each coupling constraint.

\vspace{-1mm}
\subsection{ADMM Formulation}
The ADMM \cite{Boyd2011} makes use of the decoupled structure in (\ref{eq:DecomposedModel}) by performing alternating minimizations over sets $\mathcal{\hat{X}}$ and $\mathcal{Z}$. At any iteration $k$, ADMM generates a new iterate by solving the following subproblems, until a satisfactory convergence is achieved:
%
\begin{subequations} \label{eq:ADMM}
\begin{align}
    & \boldsymbol{\hat{x}}^{k+1} \coloneqq \underset{\boldsymbol{\hat{x}} \: \in \: \mathcal{\hat{X}}}{\mbox{argmin}} \; [ F(\boldsymbol{\hat{x}},\boldsymbol{z}) + \underset{h \in \mathcal{H}}{{\sum}} L_h ], \label{eq:xk+1}
    \\
    & \boldsymbol{z}^{k+1}_{h} \coloneqq \underset{\boldsymbol{z}_{h} \: \in \: \mathcal{Z}_{h}}{\mbox{argmin}} \; [ g_h(\boldsymbol{z}_{h}) + L_h ] \hspace{8mm} \forall \: h \in \mathcal{H}, \label{eq:zk+1}
    \\
    & {\lambda}^{k+1}_{h,t} \coloneqq {\lambda}^{k}_{h,t} + \rho({\hat{p}_{h,t}}^{k+1} - {p}_{h,t}^{k+1}) \qquad \forall \; h \in \mathcal{H}, \; t \in \mathcal{T},\label{eq:Lk+1}
\end{align}
\end{subequations}
\normalsize
where (\ref{eq:xk+1}) is the subproblem solved at each step by an aggregator (holding $\boldsymbol{p}$ constant at $k$), (\ref{eq:zk+1}) denotes the subproblem of each individual household (holding $\boldsymbol{\hat{p}}$ constant at $k+1$, results of the network subproblem), and (\ref{eq:Lk+1}) is the dual update. 
Since household problems are decoupled, they can be solved in parallel. 
\vspace{-1mm}
\section{Implementation}
\subsection{Algorithm Specifications}
Primal and dual residuals are used to define the stopping criteria \cite{Mhanna2019}, which are, respectively:
\begin{subequations} \label{eq:Residuals}
\begin{align}
\boldsymbol{r}^{k} &= (\hat{p}_{h,t}^k - {p}_{h,t}^k)^{\top}, \label{eq:Residuals_primal}\\
\boldsymbol{s}^{k} &= ({p}_{h,t}^k - {p}_{h,t}^{k-1})^{\top} \label{eq:Residuals_dual},%
\end{align}
\end{subequations}
where (\ref{eq:Residuals_primal}) represent the constraint violations (i.e., (\ref{eq:Decomposed1})) at the current solution, and (\ref{eq:Residuals_dual}) represents the violation of the Karush-Kuhn-Tucker (KKT) stationarity constraints at the current iteration. The termination criteria are then given by:
\begin{equation} \label{eq:termination}
    {\lVert\boldsymbol{r}^{k}\rVert}_2 \leq \epsilon^{\text{pri}} \quad \text{and} \quad {\lVert\boldsymbol{s}^{k}\rVert}_2 \leq \epsilon^{\text{dual}},
\end{equation}
where $\epsilon^{\text{pri}}$ and $\epsilon^{\text{dual}}$ are feasibility tolerances determined by the following equations \cite{Boyd2011}:
\begin{subequations} \label{eq:Tolerances}
\begin{align}
\epsilon^{\text{pri}} &= \sqrt{H} \epsilon^{\text{abs}} + \epsilon^{\text{rel}} \text{max} \big\{ {\lVert\mathbf{\boldsymbol{\hat{p}}^{k}}\rVert}_2, {\lVert\mathbf{\boldsymbol{p}^{k}}\rVert}_2 \big\}, \label{eq:Tolerances_primal}\\
\epsilon^{\text{dual}} &= \sqrt{H} \epsilon^{\text{abs}} + \epsilon^{\text{rel}} {\lVert\mathbf{\boldsymbol{\lambda}^{k}}\rVert}_2 \label{eq:Tolerances_dual},%
\end{align}
\end{subequations}
where $\boldsymbol{\hat{p}}$ and $\boldsymbol{p}$ are vectors composed by all variables $\hat{p}_{h,t}$ and ${p}_{h,t}$ (\ref{eq:Decomposed1}), $\boldsymbol{\lambda}^{k}$ is the vector composed by all ${\lambda}_{h,t}^{k}$ (\ref{eq:Lk+1}), $\epsilon^{\text{abs}}, \epsilon^{\text{rel}} \in {\rm I\!R}_{+}$ and their values are, in turn, part of the analysis described in Section V. 
Using smaller values for these tolerances yields more accurate results. However, this requires a higher number of iterations, which directly impacts the total computation time. This may lead to inefficient tolerance values, which is investigated.
%
Finally, an adaptive residual balancing method is used to update the value of $\rho$ according to the magnitude of residuals, as described in \cite{Mhanna2019}.

\vspace{-1mm}
\subsection{Hardware description}

The aggregator subproblem (\ref{eq:xk+1}) is solved on a 32 GB RAM, Intel i7-7700, 3.60 GHz PC. Five prosumer subproblems (\ref{eq:zk+1}) are solved in parallel on five different Raspberry Pis model 3B+, 1 GB RAM, BCM2837B0, 1.4 GHz (RPis), and the remaining prosumer subproblems are solved serially on the PC. 
All problems were implemented in Python using Pyomo \cite{Pyomo2011Hart} as a modeling interface, and solved using Ipopt v3.12.11 \cite{IPOPT}, with linear solver MA27 \cite{HSL1}, in both the RPis and the PC. 
The PC is connected to the internet with a standard cable connection, and acts as a multi-client UDP server. All RPis are connected to the internet via WiFi, and act as UDP clients in an edge computing framework.

\vspace{-1mm}
\subsection{Test networks}
\begin{figure}[!t]
    \centering
    \resizebox{!}{3.5cm}{\begin{tikzpicture}

\filldraw [red] (0,0) circle (3pt);
\filldraw [black] (0.5,0.5) circle (3pt);
\filldraw [black] (1,1) circle (3pt);
\filldraw [black] (1.5,1.5) circle (3pt);
\filldraw [black] (2.3,1.5) circle (3pt);
\filldraw [black] (3.1,1.5) circle (3pt);
\draw (0.1,0.1) -- (0.5,0.5) -- (1,1) -- (1.5,1.5) -- (2.3,1.5) -- (3.1,1.5);

\filldraw [black] (1.2,0.5) circle (3pt);
\filldraw [black] (1.9,0.5) circle (3pt);
\filldraw [black] (2.4,1.0) circle (3pt);
\filldraw [black] (3.1,1.0) circle (3pt);
\filldraw [black] (3.8,1.0) circle (3pt);
\filldraw [black] (4.5,1.0) circle (3pt);
\draw (0.5,0.5) -- (1.2,0.5) -- (1.9,0.5) -- (2.4,1.0) -- (3.1,1.0) -- (3.8,1.0) -- (4.5,1.0);

\filldraw [black] (2.4,0.0) circle (3pt);
\filldraw [black] (3.1,0.0) circle (3pt);
\filldraw [black] (3.8,0.0) circle (3pt);
\filldraw [black] (4.3,-0.5) circle (3pt);
\filldraw [black] (4.8,-1.0) circle (3pt);
\filldraw [black] (4.8,-1.7) circle (3pt);
\draw (1.9,0.5) -- (2.4,0.0) -- (3.1,0.0) -- (3.8,0.0) -- (4.3,-0.5) -- (4.8,-1.0) -- (4.8,-1.7);

\filldraw [black] (2.4,-0.7) circle (3pt);
\filldraw [black] (2.4,-1.4) circle (3pt);
\filldraw [black] (1.7,-1.4) circle (3pt);
\filldraw [black] (1.0,-1.4) circle (3pt);
\draw (2.4,0.0) -- (2.4,-1.4) -- (1.7,-1.4) -- (1.0,-1.4);

\filldraw [black] (2.4,-2.1) circle (3pt);
\filldraw [black] (2.9,-2.6) circle (3pt);
\filldraw [black] (3.6,-2.6) circle (3pt);
\filldraw [black] (4.1,-3.1) circle (3pt);
\draw (2.4,-1.4) -- (2.4,-2.1) -- (2.9,-2.6) -- (3.6,-2.6) -- (4.1,-3.1);


\filldraw [black] (5.2,1.0) circle (3pt);
\filldraw [black] (5.7,0.5) circle (3pt);
\filldraw [black] (5.7,-0.2) circle (3pt);
\draw (4.5,1.0) -- (5.2,1) -- (5.7,0.5) -- (5.7,-0.2);

\filldraw [black] (5.9,1) circle (3pt);
\filldraw [black] (6.6,1) circle (3pt);
\filldraw [black] (7.1,1.5) circle (3pt);
\filldraw [black] (7.8,1.5) circle (3pt);
\filldraw [black] (8.3,2) circle (3pt);
\draw (5.2,1.0) -- (5.9,1) -- (6.6,1) -- (7.1,1.5) -- (7.8,1.5) -- (8.3,2);

\filldraw [black] (7.1,0.5) circle (3pt);
\filldraw [black] (7.8,0.5) circle (3pt);
\filldraw [black] (8.3,1) circle (3pt);
\filldraw [black] (8.8,1.5) circle (3pt);
\draw (6.6,1) -- (7.1,0.5) -- (7.8,0.5) -- (8.3,1) -- (8.8,1.5);

\filldraw [black] (8.3,0) circle (3pt);
\draw (7.8,0.5) -- (8.3,0);

\filldraw [black] (5.5,-2.3) circle (3pt);
\filldraw [black] (6,-2.8) circle (3pt);
\filldraw [black] (6.5,-3.3) circle (3pt);
\filldraw [black] (7,-2.8) circle (3pt);
\draw (4.8,-1.7) -- (5.5,-2.3) -- (6,-2.8) -- (6.5,-3.3) -- (7,-2.8);

\filldraw [black] (6,-1.7) circle (3pt);
\filldraw [black] (6.5,-1.2) circle (3pt);
\filldraw [black] (6.5,-0.5) circle (3pt);
\filldraw [black] (7.2,-0.5) circle (3pt);
\draw (5.5,-2.3) -- (6,-1.7) -- (6.5,-1.2) -- (6.5,-0.5) -- (7.2,-0.5);

\filldraw [black] (6.5,-2.3) circle (3pt);
\filldraw [black] (7,-1.7) circle (3pt);
\filldraw [black] (7.5,-1.2) circle (3pt);
\filldraw [black] (8.3,-1.2) circle (3pt);
\draw (6,-1.7) -- (6.5,-2.3) -- (7,-1.7) -- (7.5,-1.2) -- (8.3,-1.2);

\draw[dashed] (-0.5,2.5) -- (9.3,2.5) -- (9.3,-3.8) -- (-0.5,-3.8) -- (-0.5,2.5);
\draw[blue, dashed] (-0.25,2.25) -- (4.95,2.25) -- (4.95,-3.65) -- (-0.25,-3.65) -- (-0.25,2.25);


\Large
\node[blue] at (0,-3.35) {A};
\node at (9.0,-3.5) {B};

\end{tikzpicture}}
    \caption{26- and 51-bus networks showing buses, lines, and generator in red. The blue area encompasses 25 prosumers, and the black area 50 prosumers.}
    \label{fig:Network4}
    \vspace{2mm}
\end{figure}
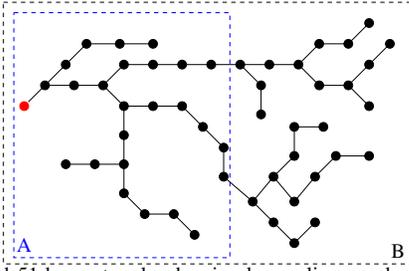

Two low-voltage distribution networks A and B, with 25 and 50 prosumers respectively, have been used for testing the proposed algorithm. They have 26 and 51 buses respectively; their configuration is illustrated in Fig. \ref{fig:Network4}. 

\begin{table}[!t]
\renewcommand{\arraystretch}{1.1}
\centering
\caption{Test cases and problem  complexity.}
\label{table:Test_cases}
\begin{tabular}{|c|c|c|c|c|}
\hline
Case & Network & $\mathcal{T}$ & No. of variables & No. of constraints\\
\hline
1 & A & $\mathcal{T}_{1}$ & 11088 & 9840\\
2 & B & $\mathcal{T}_{1}$ & 21888 & 19440\\
3 & A & $\mathcal{T}_{2}$ & 22176 & 19680\\
4 & B & $\mathcal{T}_{2}$ & 43776 & 38880\\
\hline
\end{tabular}
\end{table}

Prosumer's load and PV data used are actual power measurements, with half-hourly resolution 
on a spring day (2011/11/07), of an Australian low-voltage network. As such, we initially define $\mathcal{T}_{1} = \{0, 1, ..., 47\}, {\Delta t}_1=0.5$. Additionally, we have further split these into 15-minute resolution data sets, in which 
$\mathcal{T}_{2} = \{0, 1, ..., 95\}, {\Delta t}_2=0.25$. 

We have combined networks A and B with $\mathcal{T}_{1}$ and $\mathcal{T}_{2}$, resulting in a total of four different test cases, as seen in Table \ref{table:Test_cases}. The complexity of problem (\ref{eq:xk+1}), which takes the longest for each iteration, is also shown.



\vspace{-3mm}
\section{Results}

\begin{figure}[!t]
\centering
\subfloat{\includegraphics[scale=0.47]{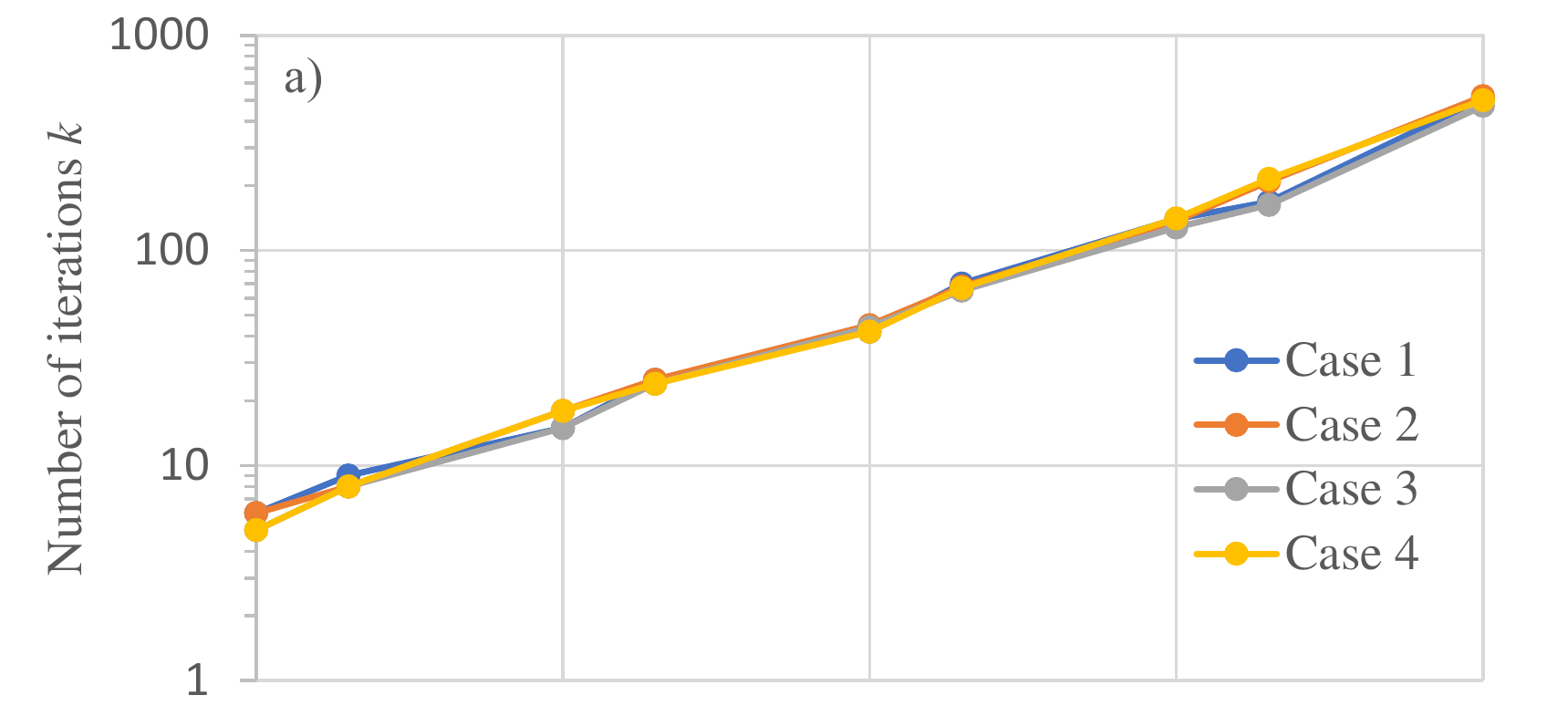} 
\label{Sample}}
\hfill
\subfloat{\includegraphics[scale=0.35]{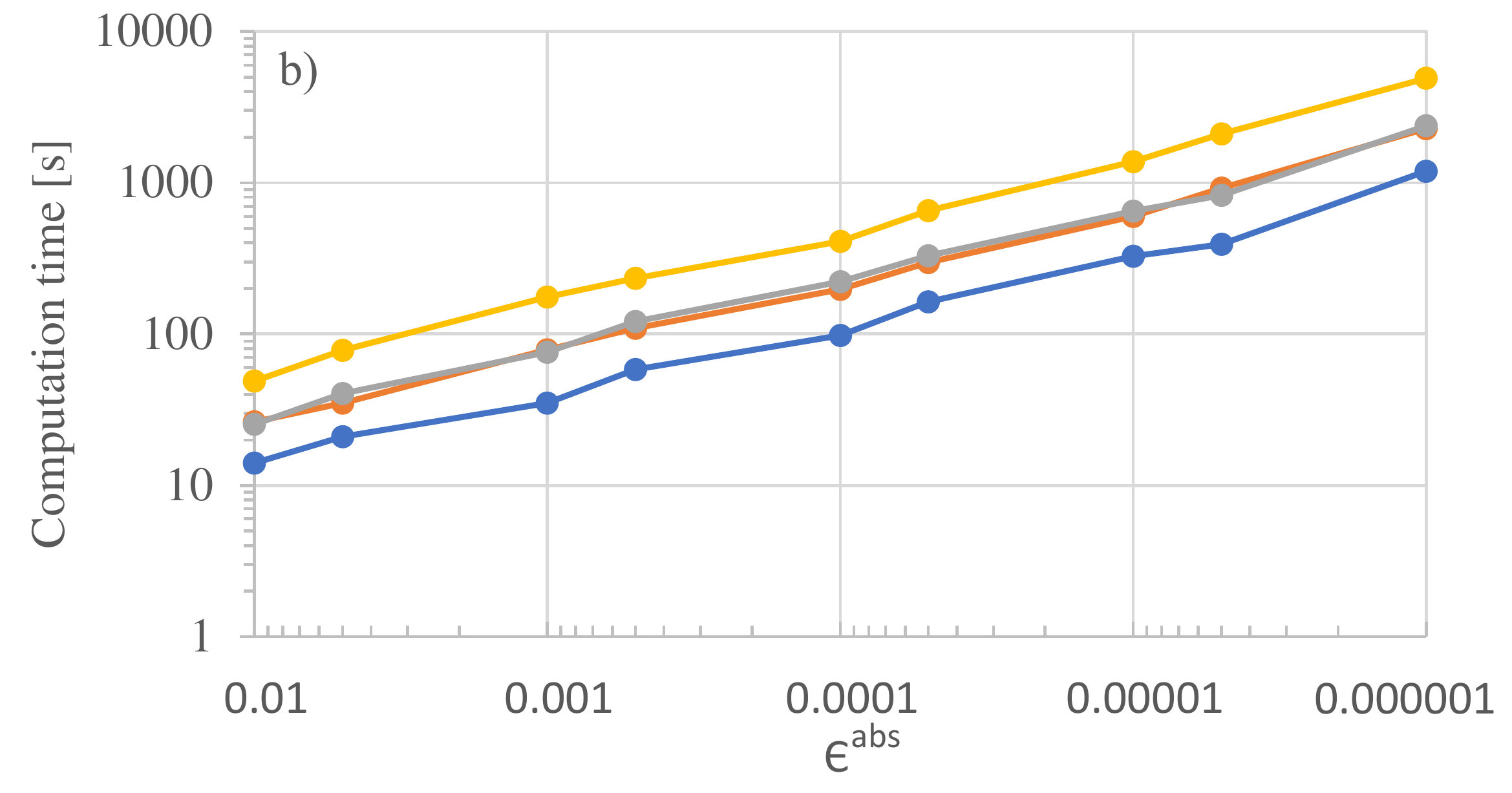}  
\label{Sample_b}}
\caption{Results for all four cases: a) depicts number of iterations $k$, and b) shows the total parallel computation time across different values for $\epsilon^{\text{abs}}$.}
\label{fig:Plot_41} 
\vspace{2mm}
\end{figure}

The results for the four test cases, with varying tolerances, are depicted on Fig. \ref{fig:Plot_41}. Throughout our tests, we have used $\epsilon^{\text{rel}} = 10 \: \epsilon^{\text{abs}}$, and $\epsilon^{\text{abs}} \in [10^{-2}, 5\times10^{-3}, 10^{-3}, ..., 5\times10^{-6}, 10^{-6}]$ for a total of nine tolerances. 

Fig. \ref{fig:Plot_41}a) shows the number of iterations $k$ each case takes to converge, across different tolerances. It is notable $k$ is very similar across all four cases, and therefore mostly independent of the problem size, which demonstrates the scalability of ADMM \cite{Mhanna2019}. 

We discuss the results in four areas, namely: computation time, system operation under congested conditions, precision of solutions, and communication requirements.

\vspace{-2mm}
\subsection{Computation Time}
%

The computation time per iteration is shown in Table \ref{table:Comp_times}. The term $t_{(\textrm{9a})}$ refers to the execution time (\ref{eq:xk+1}) in the PC, $t_{(\textrm{9b})}$ is determined by the slowest execution time of (\ref{eq:zk+1}) in the RPis, and $t_{(\textrm{9c})}$ refers to the dual update (\ref{eq:Lk+1}) execution on the PC. The average total parallel computation time per iteration is shown in the last column, $t_{\text{comp}}$, representing the time per iteration a fully distributed implementation would require.


\begin{table}[!t]
\renewcommand{\arraystretch}{1.3}
\centering
\caption{Average computation time per iteration, in seconds.}
\label{table:Comp_times}
\begin{tabular}{|c|c|c|c|}
\hline
Case & $t_{(\textrm{9a})}+t_{(\textrm{9c})} [ \SI{}{\second} ]$ & $t_{(\textrm{9b})} [ \SI{}{\second} ]$ & $t_{\text{comp}} [ \SI{}{\second} ]$\\
\hline
1 & 2.09 & 0.25 & 2.34\\
2 & 4.13 & 0.25 & 4.38\\
3 & 4.65 & 0.41 & 5.06\\
4 & 9.36 & 0.41 & 9.77\\
\hline
\end{tabular}
\end{table}


In hindsight, the solution time for the DOPF subproblem is much more predominant in the total solution time. Albeit the number of iterations $k$ remains very similar when increasing the size of the problem, the central computation time increases linearly, as seen in Table \ref{table:Comp_times} and consequently, most of the computation load in Fig. \ref{fig:Plot_41}b) stems from solving (\ref{eq:xk+1}). 

\subsection{System Operation under Congested Conditions}


\begin{figure}[!t]
\centering
\includegraphics[scale=0.325]{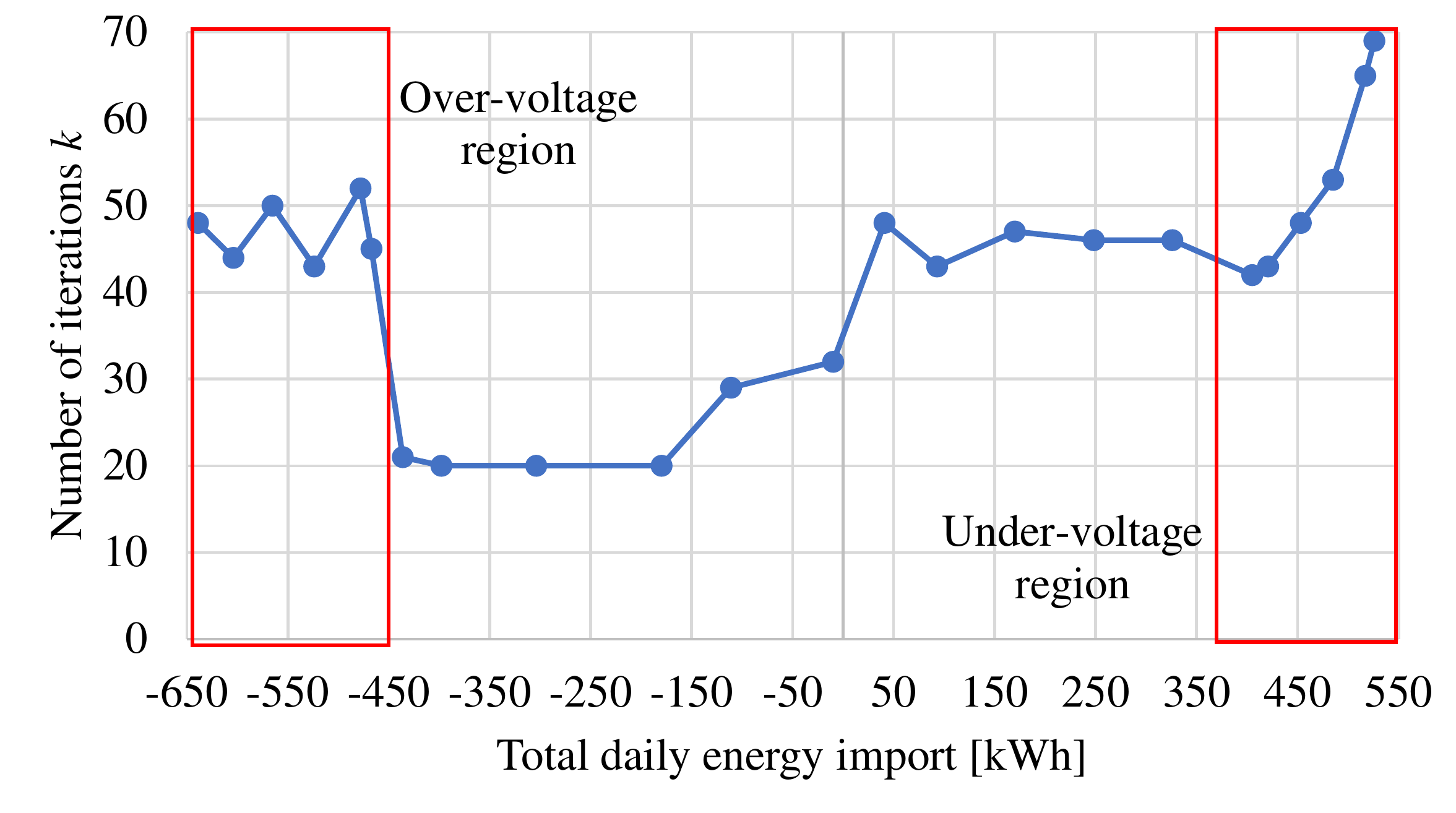}
  \caption{Number of iterations $k$ across different mixes of energy, for case 1 and $\epsilon^{\text{abs}} = 10^{-4}$.}
  \label{fig:Plot_Congested} 
\end{figure}

In real systems, demand and generation vary, which may lead to operation under congested conditions (e.g., over- or under-voltage). DOPF implementations need to be robust against these changes, even if they cause a higher number of iterations. 

To test the impact of congested conditions, demand and generation have been modified for Case 1, with a fixed tolerance of $\epsilon^{\text{abs}} = 10^{-4}$. The results in Fig. \ref{fig:Plot_Congested} show the number of iterations $k$ across different mixes of energy. The points at which constraints are active (under- and over-voltage, or input feeder limit) are also denoted in the figure, showing a clear correlation of increased $k$ on operation under congested conditions. The maximum value of $k$ does not exceed 70, roughly twice as high when compared to the average $k$ for normal operation conditions. Moreover, it is visible that a system which has surplus of energy generation converges more rapidly than a system which needs to import more energy from the upstream network. 


\subsection{Precision of Solutions}

A comparison between the optimal solution $F(\boldsymbol{x},\boldsymbol{z})$ of the central problem (\ref{eq:Central_OPF_2nd}) and each test case is shown in the third column of Table \ref{table:Tolerances}. The values demonstrate the evolution of the solution precision, showing that there is almost no variation to the end result when using very low tolerance values. Not only that, but the number of iterations to reach convergence (and consequently, the computation time) becomes prohibitive, as seen in Fig. \ref{fig:Plot_41}b).

The physical implication of different tolerances are shown in Table \ref{table:Tolerances}. It depicts the maximum ($r^{\text{max}}$) and average ($\Bar{r}$) violations of constraint (\ref{eq:Decomposed1}) - the definition of primal residual (\ref{eq:Residuals_primal}). In other words, the difference between the copies of prosumer power profiles for the network and for the household. The performance across all cases are similar even if the network sizes and $\mathcal{T}$ are different.

\begin{table}[!ht]
\renewcommand{\arraystretch}{1.1}
\centering
\caption{Solution deviation versus central optimal, maximum and average primal residuals over five different tolerances for test cases 1, 2, 3 and 4.}
\label{table:Tolerances}
\begin{tabular}{|c||c|c|c|c|}
\hline
$\epsilon^{\text{abs}}$ & Case & $F_{\%}$ & $r^{\text{max}}$ [\SI{}{\watt}] & $\Bar{r}$ [\SI{}{\watt}] \\
\hline
\multirow{4}{*}{$10^{-2}$} & 1 & +57.9 & 198.64 & 45.21 \\
& 2 & +56.2 & 260.50 & 58.89 \\
& 3 & +52.1 & 101.25 & 31.39 \\
& 4 & +61.2 & 98.12 & 38.26 \\
\hline
\multirow{4}{*}{$10^{-3}$} & 1 & +5.98 & 70.958 & 5.547 \\
& 2 & +7.42 & 33.697	& 6.174\\
& 3 & +6.65 & 10.000 & 3.032 \\
& 4 & +7.95 & 10.000 & 3.439\\
\hline
\multirow{4}{*}{$10^{-4}$} & 1 & +1.34 & 0.8082 & 0.5882\\
& 2 & +1.47 & 0.8295 & 0.6237\\
& 3 & +1.35 & 0.4813 & 0.3351 \\
& 4 & +1.50 & 1.0317 & 0.3732 \\
\hline
\multirow{4}{*}{$10^{-5}$} & 1 & +1.05 & 0.2894 & 0.0495\\
& 2 & +1.24 & 0.2088 & 0.0663\\
& 3 & +1.01 & 0.0408 & 0.0052\\
& 4 & +1.32 & 0.1290 & 0.0050 \\
\hline
\multirow{4}{*}{$10^{-6}$} & 1 & +0.99 & 0.0212 & 0.0031\\
& 2 & +1.18 & 0.0285 & 0.0043\\
& 3 & +0.97 & 0.0147 & 0.0011\\
& 4 & +1.28 & 0.0065 & 0.0011\\
\hline
\end{tabular}
\end{table}

\subsection{Communication Requirements}
The message size at each iteration between prosumers and aggregator is proportional to the choice of $\mathcal{T}$. For $\mathcal{T}_1$, the message size is smaller than 1 KB, and for $\mathcal{T}_2$ it is smaller than 2 KB. The choice of different communication protocols (UDP/TCP/HTTP) is only marginally relevant, and they are capable of dealing with these message sizes, which are much smaller than the lower limits of current mobile broadband networks download and upload speeds \cite{OpenSignal,Grigorik}. 

The actual implementation of the DOPF can utilize different structures between prosumers and the aggregator. The recent Australian trial \cite{Scott2019} has utilized an hierarchical structure where groups of prosumers send their information to local computers (Reposit boxes\footnote{https://repositpower.com/}), which then compute prosumer subproblems and communicate to a central aggregator every iteration, sending the final solution (i.e., their scheduling information) back to prosumers when the solution is achieved. However, it is possible to make full use of decentralized implementation of prosumers with edge computing hardware, as shown by the computation times of the prosumer subproblem on RPis. 

This would require communication between the aggregator and prosumers at every iteration, all of which would be located within the same geographical region (e.g., in the same low-voltage network neighborhood). The communication could be achieved, for example, with the use of \emph{last mile networks} (4G and 5G). Modern network technologies offer low latencies for this kind of application. For example, 4G network latency\footnote{We refer to \cite{OpenSignal} when defining latency as the delay between agents as data makes a round trip through the communications network.} range from \SI{30}{} to \SI{160}{\milli\second}, and upcoming 5G networks will further reduce these values \cite{OpenSignal}. In parallel, network technologies tailored for the Internet of Things \cite{Gebbran_AUPEC}, such as LTE-M, NB-IoT and EC-GSM-IoT, could also be used to deploy this communication. These networks have latencies of \SI{300}{} to \SI{600}{\milli\second} in areas within the normal cell edge of the radio cell \cite{Liberg2017cellular}. 

From the technical aspect, the solution time per iteration of the DOPF, as shown in Table \ref{table:Comp_times}, is more predominant than the latency delay of last mile networks. If implemented in a 4G network, the latency (assume an average of \SI{100}{\milli\second}) in cases 1 to 4 would take, respectively, \SI{4.3}{\percent}, \SI{2.2}{\percent}, \SI{2}{\percent} and \SI{1}{\percent} of the total time per iteration. Economical aspects could weight in more when choosing the appropriate technology to deploy this infrastructure, as well as limiting factors such as low area coverage or poor internet connection \cite{Scott2019, Guo_2018}.


\vspace{-1mm}
\section{General Comments}

The computation time of the DOPF approach grows linearly with the size of the problem, which in turn imposes a limit on the available solution time. For instance, when using a rolling horizon, the window interval for each horizon to be completed must be compatible with the DOPF solution time. For instace, larger networks with over one hundred prosumers, as simulated by the authors in \cite{Guerrero_2020}, require a longer computation time. This may not be compatible with a five-minute window interval as used by the DOPF in \cite{Scott2019}, with under fifty prosumers. 

The choice of an appropriate tolerance and time horizon $\mathcal{T}$ must take into account the problem size and the available solution time. Moreover, the communication latency and other limitations imposed by the geographical location of prosumers and the aggregator must be accounted for.
The computational burden introduced by transforming interval $\mathcal{T}_1$ into $\mathcal{T}_2$ is associated with doubling the number of variables and constraints, which in turn doubles the resolution of the problem variables.

Communication networks may not handle well the transmission of data from a very large number of prosumers to the aggregator, which happen in a very short amount of time. This may lead to congestion (data traffic above the network bandwidth) or contention (when many prosumers are trying to transmit data simultaneously) on the communication network. These problems are prone to happen when a large concentration of prosumers (over hundreds or thousands) are concentrated in the same geographical location, sharing the same communication network and a limited quantity of available resources (e.g., spectrum) from the wireless network. Nonetheless, the network latency and the message size of the communication between prosumers and aggregator are not bottlenecks when implementating the DOPF.

\vspace{-2mm}
\subsection{Future Work} 
As shown in Table \ref{table:Comp_times}, reducing the computation cost per iteration is of paramount importance for a practical implementation of the DOPF. This may include a number of strategies to reduce the computation time for each step, such as splitting (\ref{eq:xk+1}) into smaller subproblems, solved in parallel \cite{Scott2019}.

Moreover, a model to prevent the aforementioned congestion and contention problems is another suggestion for further research. This would allow for a better utilization of the available communication network resources, by allocating these resources and coordinating data transmission according to the characteristic of the DER coordination problem.

Finally, using an asynchronous ADMM may be of interest, which could improve the robustness of the algorithm against possible communication failures.

\vspace{-1mm}
\section{Conclusion}

We have implemented a DER coordination problem using DOPF, on a PC and a hardware prototype of five RPis. 
The central problem was decomposed and decoupled into a formulation suitable for solution using ADMM. 
We analyzed four different test cases, investigating the computation time and the number of iterations $k$ across different tolerances. The effect of operation under congested conditions was shown to impact $k$. We have shown trade-offs between convergence and computation speed according to solution precision. Finally, the communication requirements for the deployment of similar problems were discussed.

\vspace{-1mm}
\bibliographystyle{IEEEtran}
\bibliography{myref}{}

\end{document}